\pdfoutput=1
\RequirePackage{ifpdf}
\ifpdf 
\documentclass[pdftex]{sigma}
\else
\documentclass{sigma}
\fi

\begin{document}

\allowdisplaybreaks

\renewcommand{\PaperNumber}{045}

\FirstPageHeading

\ShortArticleName{High-Energy String Scattering Amplitudes and Signless Stirling Number Identity}

\ArticleName{High-Energy String Scattering Amplitudes\\ and Signless Stirling
Number Identity}

\Author{Jen-Chi LEE~$^\dag$, Catherine H. YAN~$^\ddag$ and Yi YANG~$^\dag$}

\AuthorNameForHeading{J.C.~Lee, C.H.~Yan and Y.~Yang}

\Address{$^\dag$~Department of Electrophysics, National Chiao-Tung University,
Hsinchu, Taiwan, R.O.C.}
\EmailD{\href{mailto:jcclee@cc.nctu.edu.tw}{jcclee@cc.nctu.edu.tw}, \href{mailto:yiyang@mail.nctu.edu.tw}{yiyang@mail.nctu.edu.tw}}

\Address{$^\ddag$~Department of Mathematics, Texas A\&M University, College Station,
TX 77843, USA}
\EmailD{\href{mailto:cyan@math.tamu.edu}{cyan@math.tamu.edu}}

\ArticleDates{Received April 23, 2012, in f\/inal form July 10, 2012; Published online July 18, 2012}

\Abstract{We give a complete proof of a set of identities
(\ref{master}) proposed recently from calculation of high-energy
string scattering amplitudes. These identities allow one to extract ratios among
high-energy string scattering amplitudes in the f\/ixed angle regime from
high-energy amplitudes in the Regge regime. The proof is based on a signless
Stirling number identity in combinatorial theory. The results are valid for
arbitrary real values $L$ rather than only for $L=0,1$ proved previously. The
identities for non-integer real value $L$ were recently shown to be realized
in high-energy compactif\/ied string scattering amplitudes [He~S., Lee~J.C., Yang~Y., arXiv:1012.3158]. The
parameter $L$ is related to the mass level of an excited string state and can
take non-integer values for Kaluza--Klein modes.}


\Classification{81T30; 83E30}



\section{Introduction}

{\sloppy Recently high-energy f\/ixed angle string scattering amplitudes were intensively
investigated \mbox{\cite{PRL,ChanLee2,Dscatt}} for string states at arbitrary mass
levels. One of the motivation of this calculation has been to uncover the
fundamental hidden stringy spacetime symmetry conjectured more than twenty
years ago in \cite{ Gross, Gross1, GrossManes, GM1, GM}. It was conjectured in late 80s
\cite{Gross,Gross1} that there exist li\-near relations or symmetries among scattering
amplitudes in the high energy f\/ixed angle regime, or Gross regime (GR). In the
recent calculations, an inf\/inite number of linear relations among high-energy
scattering amplitudes of dif\/ferent string states were derived and the complete
ratios among the amplitudes at each f\/ixed mass level can be determined. An
important new ingredient of this string amplitude calculation was based on an
old conjecture of \cite{ZNS2-2,ZNS3,ZNS2,ZNS2-1,ZNS1-1, ZNS1} on the decoupling of zero-norm states
(ZNS) in the spectrum, in particular, the identif\/ication of inter-particle
symmetries induced by the inter-particle ZNS \cite{ZNS1-1, ZNS1} in the spectrum.

}

Another fundamental regime of high-energy string scattering amplitudes is the
Regge regime (RR) \cite{RR3,RR2,RR1,RR6,RR5,RR4}. See also \cite{OA,DL,KP}. It
was found \cite{bosonic} that the high-energy string scattering amplitudes in
the GR and RR contain information complementary to each other. On the other
hand, since the decoupling of ZNS applies to all kinematic regimes, one
expects that the ratios obtained from the decoupling of ZNS in the GR are
closely related to the decoupling of ZNS or scattering amplitudes in the RR.
Moreover, it is conceivable that there exists some link between the patterns
of the high-energy scattering amplitudes in the GR and RR. It was found that
the number of high-energy scattering amplitudes for each f\/ixed mass level in
the RR is much more numerous than that of GR calculated previously. In
contrast to the case of scattering amplitudes in the
GR, there is no linear relation
among scattering amplitudes in the RR. Moreover, it was discovered that the leading
order amplitudes at each f\/ixed mass level in the RR can be expressed in terms
of the Kummer function of the second kind. More surprisingly, for those
leading order string tree four-point high-energy amplitudes $A^{(N,2m,q)}$
(see equation~(\ref{A})) in the RR with the same type of $(N,2m,q)$ as those of GR,
one can extract from them the ratios $T^{(N,2m,q)}/T^{(N,0,0)}$ (see
equations~(\ref{2}) and~(\ref{3})) in the GR by using this Kummer function. The
calculation was based on a set of identities equation~(\ref{master}) which depend on
an \textit{integer} parameter $L(M_{2}^{2})=1-N$ where $M_{2}^{2}=2(N-1)$ is
the mass square of the second string scattering state (here, for simplicity,
one chooses the other three string states to be tachyons). The calculation can
be done for both the case of open string and the closed string as well.

The proof of these identities for $L=0,1$ was previously given in
\cite{RRsusy, bosonic} based on a set of signed Stirling number identities
developed in 2007~\cite{MK}. However, the proof of these identities for
arbitrary integer values $L$ is still lacking, and it is crucial to complete
the proof in order to link high-energy string scattering amplitudes in the RR
and GR regimes as claimed above.  Moreover, it was discovered recently~\cite{HLY} that in order to link the RR and GR string amplitudes for
the scattering \textit{compactified} on tori, one needs to prove the identities
for arbitrary \textit{real} values~$L$.

In this letter, we are going to prove these identities for arbitrary real
values $L$ by using a~signless Stirling number identity. It is remarkable to
see that the identities suggested by string theory calculation can be
rigorously proved by a totally dif\/ferent mathematical method in combinatorial
theory. It is also very interesting to see that, physically, the identities
for arbitrary \textit{real} values $L$ in equation~(\ref{L}) can only be realized in
high-energy compactif\/ied string scattering
amplitudes considered very recently~\cite{HLY}. This is mainly due to the relation $M^{2}=\left(  K^{25}\right)
^{2}+\hat{M}^{2}$ where $\hat{M}^{2}=2(N-1)$ and $K^{25}=\frac{2\pi
l-\theta_{l}+\theta_{i}}{2\pi R}$~(see, e.g.,~\cite{POL}) is the generalized KK internal
momentum corresponding to the compactif\/ied string coordinate~\cite{HLY}. In
the def\/inition of $K^{25}$, $l$ is the quantized momentum, $R$ is the radius
of compactif\/ied $S^{1}$, and we have included a~nontrivial Wilson line with
$U(n)$ Chan--Paton factors, $i,l=1,2,\dots,n.$ All other high-energy string
scattering amplitudes calculated previously~\cite{RRsusy, bosonic} correspond
to integer values of~$L$ only.

\section{GR and RR amplitudes}

We begin with a brief review of the
high-energy string scattering in the f\/ixed
angle regime,
\begin{gather*}
s,-t\rightarrow\infty,\qquad t/s\approx-\sin^{2}\frac{\phi}{2}=\text{f\/ixed} \quad (\text{but} \ \phi\neq0), 
\end{gather*}
where $s$, $t$ and~$u$ are the Mandelstam variables and $\phi$ is the CM
scattering angle. It was shown~\cite{PRL} that for the 26D open bosonic string
the only states that will survive the high-energy limit at mass level
$M_{2}^{2}=2(N-1)$ are of the form (we choose the second state of the
four-point function to be the higher spin string state)
\begin{gather}
\left\vert N,2m,q\right\rangle \equiv\big(\alpha_{-1}^{T}\big)^{N-2m-2q}\big(\alpha
_{-1}^{L}\big)^{2m}\big(\alpha_{-2}^{L}\big)^{q}|0,k\rangle, \label{2}
\end{gather}
where the polarizations of the 2nd particle with momentum $k_{2}$ on the
scattering plane were def\/ined to be $e^{P}=\frac{1}{M_{2}}(E_{2}, k_{2},0)=\frac{k_{2}}{M_{2}}$ as the momentum polarization,
$e^{L}=\frac{1}{M_{2}}(k_{2},E_{2},0)$ the longitudinal polarization
and $e^{T}=(0,0,1)$ the transverse polarization which lies on the scattering
plane. The three vectors $e^{P}$, $e^{L}$ and $e^{T}$ satisfy the completeness
relation~\cite{ChanLee2} $\eta_{\mu\nu}=\sum_{\alpha,\beta}e_{\mu}^{\alpha
}e_{\nu}^{\beta}\eta_{\alpha\beta}$ where $\mu,\nu=0,1,2$ and $\alpha
,\beta=P,L,T$, and $\alpha_{-1}^{T}=\sum_{\mu}e_{\mu}^{T}\alpha_{-1}^{\mu}$,
$\alpha_{-1}^{T}\alpha_{-2}^{L}=\sum_{\mu}e_{\mu}^{T}e_{\nu}^{L}\alpha
_{-1}^{\mu}\alpha_{-2}^{\nu}$ etc. ${\rm diag}\,\eta_{\mu\nu}=(-1,1,1)$. In
equation~(\ref{2}), $N$, $m$ and $q$ are non-negative integers and $N\geq2m+2q.$ These
integers characterise the mass square and ``spin'' of the higher string states.
Note that~$e^{P}$ approaches to~$e^{L}$ in the GR~\cite{ChanLee2} or
equivalently the $e^{P}$ polarizations can be gauged away using ZNS. So we did
not put~$e^{P}$ components in equation~(\ref{2}). For simplicity, we choose the
particles associated with momenta $k_{1}$, $k_{3}$ and $k_{4}$ to be tachyons.
It turned out that the high-energy f\/ixed angle scattering amplitudes can be
calculated by using the saddle-point method~\cite{PRL}. An inf\/inite number of
linear relations among four-point high-energy scattering amplitudes of
dif\/ferent string states were derived and the complete ratios among the
amplitudes at each f\/ixed mass level can be calculated to be~\cite{PRL}
\begin{gather}
\frac{T^{(N,2m,q)}}{T^{(N,0,0)}}=\left(  -\frac{1}{M_{2}}\right)
^{2m+q}\left(  \frac{1}{2}\right)  ^{m+q}(2m-1)!!. \label{3}
\end{gather}
Alternatively, the ratios can be calculated by the method of decoupling of two
types of ZNS in the old covariant f\/irst quantized string spectrum. Since the
decoupling of ZNS applies to all string loop order, the ratios calculated in
equation~(\ref{3}) are valid to all string loop order. Similarly, the ratios for
closed string, superstring and D-brane scattering amplitudes~\cite{Dscatt} can be obtained.

Another high-energy regime of string scattering amplitudes, which
contains complementary information of the theory, is the f\/ixed momentum
transfer $t$ or RR. That is in the kinematic regime%
\begin{gather*}
s\rightarrow\infty,\qquad \sqrt{-t}=\text{f\/ixed}\quad (\text{but}\ \sqrt{-t}\neq\infty).
\end{gather*}
It was found \cite{bosonic} that the number of high-energy scattering
amplitudes for each f\/ixed mass level in this regime is much more numerous than
that of f\/ixed angle regime calculated previously. On the other hand, it seems
that both the saddle-point method and the method of decoupling of zero-norm
states adopted in the calculation of f\/ixed angle regime do not apply to the
case of Regge regime. However the calculation is still manageable, and the
general formula for the high energy $(s,t)$ channel open string scattering
amplitudes at each f\/ixed mass level can be written down explicitly.

It was shown that a class of high-energy open string states in the Regge
regime at each f\/ixed mass level $N=\sum_{l,m}lp_{l}+mq_{m}$ are
\cite{RRsusy, bosonic}
\begin{gather}
\left\vert p_{l},q_{m}\right\rangle =\prod_{l>0}\big(\alpha_{-l}^{T}\big)^{p_{l}}
\prod_{m>0}\big(\alpha_{-m}^{L}\big)^{q_{m}}|0,k\rangle. \label{RR}
\end{gather}
As explained in \cite{RRsusy} for the purpose of connecting the RR with the GR
limit of a scattering amplitude, suf\/f\/ices to consider scattering amplitudes
involving only the vertex in equation~(\ref{RR}). The complete high energy vertex
can be found in~\cite{RRsusy}. The momenta of the four particles on the
scattering plane are
\begin{alignat*}{3}
& k_{1}  =\left(  +\sqrt{p^{2}+M_{1}^{2}},-p,0\right)  ,\qquad &&
k_{2}     =\left(  +\sqrt{p^{2}+M_{2}^{2}},+p,0\right)  ,&& \\
& k_{3}  =\left(  -\sqrt{q^{2}+M_{3}^{2}},-q\cos\phi,-q\sin\phi\right)  ,\qquad &&
k_{4}  =\left(  -\sqrt{q^{2}+M_{4}^{2}},+q\cos\phi,+q\sin\phi\right),&
\end{alignat*}
where $p\equiv\left\vert \vec{p} \right\vert $, $q\equiv\left\vert
 \vec{q} \right\vert $ and $k_{i}^{2}=-M_{i}^{2}$. The relevant
kinematics are
\begin{gather*}
e^{P}\cdot k_{1}\simeq-\frac{s}{2M_{2}},\qquad e^{P}\cdot k_{3}\simeq
-\frac{\tilde{t}}{2M_{2}}=-\frac{t-M_{2}^{2}-M_{3}^{2}}{2M_{2}},
\\
e^{L}\cdot k_{1}\simeq-\frac{s}{2M_{2}},\qquad e^{L}\cdot k_{3}\simeq
-\frac{\tilde{t}^{\prime}}{2M_{2}}=-\frac{t+M_{2}^{2}-M_{3}^{2}}{2M_{2}},
\end{gather*}
and
\begin{gather*}
e^{T}\cdot k_{1}=0, \qquad e^{T}\cdot k_{3}\simeq-\sqrt{-{t}},
\end{gather*}
where $\tilde{t}$ and $\tilde{t}^{\prime}$ are related to $t$ by f\/inite mass
square terms
\[
\tilde{t}=t-M_{2}^{2}-M_{3}^{2}, \qquad \tilde{t}^{\prime}=t+M_{2}%
^{2}-M_{3}^{2}.
\]
Note that $e^{P}$ does not approach to $e^{L}$ in the RR. The Regge scattering
amplitude for the $(s,t)$ channel was calculated to be~\cite{bosonic} (we
choose to calculate $e^{L}$ amplitudes, the $e^{P}$ amplitudes can be
similarly discussed)
\begin{gather}
A^{(N,2m,q)}(s,t)=s^{\alpha(t)}\sqrt{-t}^{N-2m-2q}\left(  \frac{1}{2M_{2}%
}\right)  ^{2m+q}\cdot2^{2m}(\tilde{t}^{\prime})^{q}U\left(  -2m , \frac
{t}{2}+2-2m , \frac{\tilde{t}^{\prime}}{2}\right),  \label{A}
\end{gather}
where the level independent \cite{bosonic} exponent $\alpha(t)=a(0)+\alpha
^{\prime}t$,  $a(0)=1$ and $\alpha^{\prime}=\frac{1}{2}$. In equation~(\ref{A}) $U$
is the Kummer function of the second kind and is def\/ined to be
\begin{gather*}
U(a,c,x)=\frac{\pi}{\sin\pi c}\left[  \frac{M(a,c,x)}{(a-c)!(c-1)!}%
-\frac{x^{1-c}M(a+1-c,2-c,x)}{(a-1)!(1-c)!}\right],  \qquad  c\neq 2,3,4,\dots,
\end{gather*}
where $M(a,c,x)=\sum\limits_{j=0}^{\infty}\frac{(a)_{j}}{(c)_{j}}\frac{x^{j}}{j!}$ is
the Kummer function of the f\/irst kind. Note that the second argument of Kummer
function $c=\frac{t}{2}+2-2m$ is a function of the variable $t$, and is not a~constant as it was in the literature previously.

It can be seen from equation~(\ref{A}) that the Regge scattering amplitudes at each
f\/ixed mass level are no longer proportional to each other. The ratios are $t$
dependent functions and can be calculated to be \cite{bosonic}
\begin{gather}
\frac{A^{(N,2m,q)}(s,t)}{A^{(N,0,0)}(s,t)}    =(-1)^{m}\left(  -\frac
{1}{2M_{2}}\right)  ^{2m+q}(\tilde{t}^{\prime}-2N)^{-m-q}(\tilde{t}^{\prime
})^{2m+q}\nonumber\\
\hphantom{\frac{A^{(N,2m,q)}(s,t)}{A^{(N,0,0)}(s,t)}    =}{}
\times  \sum_{j=0}^{2m}(-2m)_{j}\left(  -1+N-\frac{\tilde{t}^{\prime}}%
{2}\right)  _{j}\frac{(-2/\tilde{t}^{\prime})^{j}}{j!}+\mathit{O}\left\{
\left(  \frac{1}{\tilde{t}^{\prime}}\right)  ^{m+1}\right\},  \label{Ratio}
\end{gather}
where $(x)_{j}=x(x+1)(x+2)\cdots (x+j-1)$ is the Pochhammer symbol which can be
expressed in terms of the signed Stirling number of the f\/irst kind $s\left(
n,k\right)  $ as following%
\begin{gather*}
\left(  x\right)  _{n}=\sum_{k=0}^{n}(-)^{n-k}s\left(  n,k\right)  x^{k}.
\end{gather*}
It was proposed in \cite{bosonic} that the coef\/f\/icients of the leading power
of $\tilde{t}^{\prime}$ in equation~(\ref{Ratio}) can be identif\/ied with the ratios
in equations~(\ref{3}). To ensure this identif\/ication%
\begin{gather}
\lim_{\tilde{t}^{\prime}\rightarrow\infty}\frac{A^{(N,2m,q)}}{A^{(N,0,0,)}%
}=\frac{T^{(N,2m,q)}}{T^{(N,0,0)}}=\left(  -\frac{1}{M_{2}}\right)
^{2m+q}\left(  \frac{1}{2}\right)  ^{m+q}(2m-1)!!,\label{Ratios2}%
\end{gather}
one needs the following identity
\begin{gather}
   \sum_{j=0}^{2m}(-2m)_{j}\left(  -L-\frac{\tilde{t}^{\prime}}{2}\right)
_{j}\frac{(-2/\tilde{t}^{\prime})^{j}}{j!}\nonumber\\
\qquad {} =0\cdot (-\tilde{t}^{\prime})^{0}\!+0\cdot (-\tilde{t}^{\prime})^{-1}\!+\dots +0\cdot(-\tilde
{t}^{\prime})^{-m+1}\!+\frac{(2m)!}{m!}(-\tilde{t}^{\prime})^{-m}+\mathit{O}
\left\{  \left(  \frac{1}{\tilde{t}^{\prime}}\right)  ^{m+1}\right\}\!,
\label{master}
\end{gather}
where $L=1-N$ and is an integer. Similar identif\/ication can be extended to the
case of closed string as well. For all four classes of high-energy superstring
scattering amplitudes, $L$ is an integer too~\cite{RRsusy}. A recent work on
string D-particle scattering amplitudes \cite{LMY} also gives an integer value of~$L$.
Note that $L$ af\/fects only the sub-leading terms in $\mathit{O}\Big\{
\left(  \frac{1}{\tilde{t}^{\prime}}\right)  ^{m+1}\Big\}$. Here we give a
simple example for $m=3$ \cite{RRsusy}
\begin{gather*}
  \sum_{j=0}^{6}(-2m)_{j}\left(  -L-\frac{\tilde{t}^{\prime}}{2}\right)
_{j}\frac{(-2/\tilde{t}^{\prime})^{j}}{j!}\\
\qquad{}   =\frac{120}{(-\tilde{t}^{\prime})^{3}}+\frac{720L^{2}-2640L+2080}%
{(-\tilde{t}^{\prime})^{4}}+\frac{480L^{4}-4160L^{3}+12000L^{2}-12928L+3840}%
{(-\tilde{t}^{\prime})^{5}}\\
 \quad\qquad{}  +\frac{64L^{6}-960L^{5}+5440L^{4}-14400L^{3}+17536L^{2}-7680L}{(-\tilde
{t}^{\prime})^{6}}.
\end{gather*}
Mathematically, equation~(\ref{master}) was exactly proved \cite{RRsusy, bosonic} for
$L=0,1$ by a calculation based on a set of signed Stirling number identities
developed very recently in combinatorial theory in~\cite{MK}. For general
integer $L$ cases, only the identity corresponding to the nontrivial leading
term $\frac{(2m)!}{m!}(-\tilde{t}^{\prime})^{-m}$ was rigorously proved
\cite{RRsusy}, but not for other ``0~identities''. A numerical proof of
equation~(\ref{master}) was given in~\cite{RRsusy} for arbitrary real values~$L$ and
for non-negative integer~$m$ up to $m=10.$ It was then conjectured that~\cite{RRsusy} equation~(\ref{master}) is valid for any \textit{real} number~$L$ and
any non-negative integer~$m$. It is important to prove equation~(\ref{master}) for
any non-negative integer~$m$ and arbitrary real values~$L$, since these values
can be realized in the high-energy scattering of \textit{compactified} string
states, as was shown recently in~\cite{HLY}. Real values of $L$ appear in
string compactif\/ications due to the dependence on the generalized KK internal
momenta $K_{i}^{25}$ \cite{HLY}%
\begin{gather}
L=1-N-\big(K_{2}^{25}\big)^{2}+K_{2}^{25}K_{3}^{25}.\label{L}
\end{gather}
All other high-energy string scattering amplitudes calculated previously
\cite{RRsusy, bosonic} correspond to integer value of~$L$ only. It is thus of
importance to rigorously prove the validity of equation~(\ref{master}) for arbitrary
real values~$L$.

\section{Proof of the identity}

We now proceed to prove equation~(\ref{master}). We f\/irst rewrite the left-hand side
of equation~(\ref{master}) in the following form%
\begin{gather}
\sum_{j=0}^{2m}(-2m)_{j}\left(  -L-\frac{\tilde{t}^{\prime}}{2}\right)
_{j}\frac{(-2/\tilde{t}^{\prime})^{j}}{j!}\nonumber\\
\qquad{}{} =\sum_{j=0}^{2m}\left(  -1\right)
^{j}\binom{2m}{j}\sum_{l=0}^{j}\binom{j}{l}(-L)_{j-l}\sum_{s=0}^{l}c\left(
l,s\right)  \left(  -\dfrac{2}{\tilde{t}^{\prime}}\right)  ^{j-s},
\label{Stirling expansion}
\end{gather}
where we have used the identity $(a+b)_{j}=\sum\limits_{l=0}^{j}\binom{j}{l}%
(a)_{j-l}(b)_{l}$ and have introduced the signless Stirling number of the
f\/irst kind $c\left(  l,s\right)  $ to expand the Pochhammer symbol%
\begin{gather}
\left(  x\right)  _{n}=\sum_{k=0}^{n}c\left(  n,k\right)  x^{k}.
\label{Stirling}
\end{gather}
The coef\/f\/icient of $(-2/\tilde{t}^{\prime})^{i}$ in
equation~(\ref{Stirling expansion}), which will be def\/ined as $G\left(  m,i\right)
$, can be read of\/f from the equation as%
\begin{gather*}
G\left(  m,i\right)  =\sum_{j=0}^{2m}\sum_{l=0}^{j}\left(  -1\right)
^{j+i}\binom{2m}{j}\binom{j}{l}(-L)_{j-l}c\left(  l,j-i\right)  .
\end{gather*}
One needs to prove that
\begin{alignat}{3}
& \text{1.}  \ & & G\left(  m,m\right)  =\left(  2m-1\right)  !!, \ \text{for all} \ L\in
\mathbb{R}; & \label{A1}\\
& \text{2.}  \ & &  G\left(  m,i\right)  =0, \ \text{for all} \ L\in
\mathbb{R} \ \text{and} \ 0\leq i<m. \label{A2}
\end{alignat}
From the def\/inition of $c\left(  n,k\right)  $ in (\ref{Stirling}), we note
that $c\left(  n,k\right)  \neq0$ only if $0\leq k\leq n$. Thus $c\left(
l,j-i\right)  \neq0$ only if $j\geq i$ and $l\geq j-i$. We can rewrite
$G\left(  m,i\right)  $ as
\begin{gather}
G\left(  m,i\right)      =\sum_{j=i}^{2m}\sum_{l=j-i}^{j}\left(  -1\right)
^{j}\binom{2m}{j}\binom{j}{l}(-L)_{j-l}c\left(  l,j-i\right) \nonumber\\
\hphantom{G\left(  m,i\right)}{}
 =\sum_{k=0}^{2m-i}\sum_{p=0}^{i}\left(  -1\right)  ^{k+i}\binom{2m}%
{i+k}\binom{i+k}{p+k}(-L)_{i-p}c\left(  k+p,k\right) \nonumber\\
\hphantom{G\left(  m,i\right)}{}  =\left(  -1\right)  ^{i}\sum_{p=0}^{i}(-L)_{i-p}\binom{2m}{i-p}\sum
_{k=0}^{2m-i}\left(  -1\right)  ^{k}\binom{2m-i+p}{k+p}c\left(  k+p,k\right)
\nonumber\\
\hphantom{G\left(  m,i\right)}{} \equiv\left(  -1\right)  ^{i}\sum_{p=0}^{i}(-L)_{i-p}\binom{2m}
{i-p}S_{2m-i}\left(  p\right),  \label{G(m,i)}
\end{gather}
where we have def\/ined
\begin{gather}
S_{N}\left(  p\right)  =\sum_{k=0}^{N}\left(  -1\right)  ^{k}\binom{N+p}%
{k+p}c\left(  k+p,k\right)  . \label{SN}
\end{gather}
It is easy to see that for f\/ixed $m$ and $0\leq i<m$, $G\left(  m,i\right)  $
is a polynomial of $L$ of degree $i$, expanded with the basis $1$, $\left(
-L\right)  _{1}$, $\left(  -L\right)  _{2}$, \dots. Note that $p\leq i<m$, so
$2m-i\geq p+1$. For equation~(\ref{A2}), we want to show that $S_{N}\left(
p\right)  =0$ for $N\geq p+1$. For this purpose, we def\/ine the functions%
\begin{gather*}
C_{n}\left(  x\right)  =\sum_{k\geq0}c\left(  k+n,k\right)  x^{k+n}.
\end{gather*}
The recurrence of the signless Stirling number identity
\begin{gather*}
c\left(  k+n,k\right)  =\left(  n+k-1\right)  c\left(  n+k-1,k\right)
+c\left(  n+k-1,k-1\right)
\end{gather*}
leads to the equation
\begin{gather*}
C_{n}\left(  x\right)  =\dfrac{x^{2}}{1-x}\dfrac{d}{dx}C_{n-1}\left(
x\right)  ,
\end{gather*}
with the initial value
\begin{gather*}
C_{0}\left(  x\right)  =\dfrac{1}{1-x}.
\end{gather*}
The f\/irst couple of $C_{n}\left(  x\right)  $\ can be calculated to be
\begin{gather*}
C_{1}\left(  x\right)  =\dfrac{x^{2}}{\left(  1-x\right)  ^{3}},\qquad C_{2}\left(  x\right)  =\dfrac{x^{4}+2x^{3}}{\left(  1-x\right)  ^{5}},
\qquad C_{3}\left(  x\right)  =\dfrac{x^{6}+8x^{5}+6x^{4}}{\left(
1-x\right)  ^{7}}.
\end{gather*}
Now by induction, it is easy to show that
\begin{gather}
C_{n}\left(  x\right)  =\dfrac{f_{n}\left(  x\right)  }{\left(  1-x\right)
^{2n+1}},\qquad \text{where} \quad f_{n}\left(  x\right)  =x^{2n}+\mathcal{O}\left(
x^{2n-1}\right)  , \label{Cn}
\\
f_{n}\left(  1\right)  =\left(  2n-1\right)  !!. \label{fn}
\end{gather}

\begin{proof}
If
\[
f_{n}\left(  1\right)  =\left(  2n-1\right)  !!,
\]
then
\begin{gather*}
f_{n+1}\left(  x\right)      =\left(  1-x\right)  ^{2n+3}\dfrac{x^{2}}
{1-x}\dfrac{d}{dx}\left[  \dfrac{f_{n}\left(  x\right)  }{\left(  1-x\right)
^{2n+1}}\right]
   =x^{2}\left[  \left(  2n+1\right)  f_{n}\left(  x\right)  +\left(
1-x\right)  f_{n}^{\prime}\left(  x\right)  \right]
\end{gather*}
when $x=1$, since $f_{n}\left(  x\right)  $\ is a polynomial, this gives%
\begin{gather*}
f_{n+1}\left(  1\right)  =\left(  2n+1\right)  f_{n}\left(  1\right)  =\left(
2n+1\right)  !!=\left[  2\left(  n+1\right)  -1\right]  !!.\tag*{\qed}
\end{gather*}\renewcommand{\qed}{}
\end{proof}

In order to prove equation~(\ref{A2}), we note that%
\begin{gather*}
(-1)^{N}S_{N}\left(  p\right)  =\sum_{k=0}^{N}\left(  -1\right)  ^{N+k}%
\binom{N+p}{k+p}c\left(  k+p,k\right)
\end{gather*}
is the coef\/f\/icient of $x^{N+p}$\ in the function
\begin{gather*}
\left(  1-x\right)  ^{N+p}C_{p}\left(  x\right)      =\left(  1-x\right)
^{N+p}\sum_{k\geq0}c\left(  k+p,k\right)  x^{k+p}\nonumber\\
\hphantom{\left(  1-x\right)  ^{N+p}C_{p}\left(  x\right) }{}
  =\sum_{k\geq0}\sum_{m=0}^{N+p}\binom{N+p}{m}\left(  -1\right)
^{N+p-m}c\left(  k+p,k\right)  x^{N+2p+k-m}.
\end{gather*}
On the other hand, the above function is a polynomial when $N>p$ due to
equation~(\ref{Cn})
\begin{gather}
\left(  1-x\right)  ^{N+p}C_{p}\left(  x\right)  =f_{p}\left(  x\right)
\left(  1-x\right)  ^{N-p-1}=x^{N+p-1}+\mathcal{O}\left(  \cdots\right)  .
\label{Cp}
\end{gather}
It is obvious that the coef\/f\/icient of $x^{N+p}$\ in equation~(\ref{Cp}) is zero.
This proves that $S_{N}\left(  p\right)  =0$ for $N>p$. For $m>i\geq0$ and
$0\leq p\leq i$, $2m-i>i\geq p$, we have $S_{2m-i}\left(  p\right)  =0$ and
thus equation~(\ref{A2}),
\begin{gather*}
G\left(  m,i\right)  =\left(  -1\right)  ^{i}\sum_{p=0}^{i}(-L)_{i-p}
\binom{2m}{i-p}S_{2m-i}\left(  p\right)  =0.
\end{gather*}

In order to prove the f\/irst identity in equation~(\ref{A1}), we f\/irst note that the
above argument\ remains true for $i=m$ and $0\leq p<i$, i.e.\ $2m-i=i>p$ and
thus $S_{2m-i}\left(  p\right)  =0$. So $G\left(  m,m\right)  $ only receives
contributions from the term with $p=i=m$. By using equation~(\ref{G(m,i)}) and~(\ref{SN}), we can evaluate
\begin{gather}
G\left(  m,m\right)  =\left(  -1\right)  ^{m}S_{m}\left(  m\right)
=\sum_{k=0}^{m}\left(  -1\right)  ^{k+m}\binom{2m}{k+m}c\left(  k+p,k\right).
\label{GMM}
\end{gather}
Equation~(\ref{GMM}) corresponds to the coef\/f\/icient of $x^{2m}$ in the
function
\begin{gather*}
(1-x)^{2m}C_{m}(x)=\dfrac{f_{m}\left(  x\right)  }{1-x}=f_{m}\left(  x\right)
\big(1+x+x^{2}+\cdots\big).
\end{gather*}
By equation~(\ref{fn}), this coef\/f\/icient is
\begin{gather*}
f_{m}\left(  1\right)  =\left(  2m-1\right)  !!.
\end{gather*}
This proves equation~(\ref{A1}). We thus have completed the proof of
equation~(\ref{master}) for any non-negative integer~$m$ and any real value~$L$.

\subsection*{Acknowledgments}
 We thank Rong-Shing Chang, Song He, Yoshihiro Mitsuka and
Keijiro Takahashi for helpful discussions. This work is supported in part by
the National Science Council, 50 billions project of Ministry of Education and
National Center for Theoretical Science, Taiwan.

\pdfbookmark[1]{References}{ref}
\LastPageEnding

\end{document}